\documentstyle[12pt]{article}
\def\rsim{\mathrel{\raise2pt\hbox to 8pt{\raise -5pt\hbox{$\sim$}\hss{$>$}}}}
\def\lsim{\mathrel{\raise2pt\hbox to 8pt{\raise -5pt\hbox{$\sim$}\hss{$<$}}}}

\begin{document}

\begin{center}
{\large

FEW-BODY PHYSICS -- THEN AND NOW}\\
\vspace*{.20in}
J. L. Friar\\
Theoretical Division\\
Los Alamos National Laboratory\\
Los Alamos, NM  87545\\
\vspace*{.20in}
{\large
ABSTRACT}
\end{center}

A summary of the XIV\underline{th} International Conference on Few-body
Problems In Physics is given, with an emphasis on the important problems solved
recently and the prognosis for the future of the field. Personal remarks and
``homework'' problem assignments are made.

\vspace*{.20in}

\begin{center}
INTRODUCTION\\
\end{center}

It is within the purview of a summary speaker (or perhaps it is even an
obligation) to reminisce about the past, and I will do so.  As a graduate
student in the middle sixties, my activities were primarily directed at nuclear
structure and electromagnetic interactions.  Nuclear structure calculations
deal with degrees of freedom in an economical way, ignoring those that don't
actively participate.  When I finished my thesis and went to CERN, I discovered
a completely different side to nuclear physics. At that time the field that we
now call medium-energy physics was forming. It was largely an amalgam of
electron scattering, high-energy (for that time) proton-nucleus scattering, and
meson-nucleus interactions; I found the mix fascinating (and still do).  I soon
discovered a set of summer school lectures by Colin Wilkin[1],  which played a
very large role in my intellectual development.  These pedagogical lectures
treated proton-nucleus scattering in the Glauber formalism.  To my delight
\underline{all} of the coordinates of the nucleus entered into his treatment,
which included momentum conservation and redundant coordinates, in a completely
microscopic approach.  My first paper as a fresh Ph.D. used information from
elastic electron scattering by $^4$He to constrain the latter's wave function,
and then applied this to p-$^4$He scattering. There were technical problems
with this work and it was never published, but I applied the same ideas to the
Coulomb energy of $^3$He, and the resulting ``hyperspherical'' approximation[2]
(derived independently by Fabre de la Ripelle) is both accurate and has
withstood the test of time.\\ \indent As a theorist, the beauty of this corner
of nuclear physics lay in my control over all aspects of a system:  the
coordinates, spins, and isospins. Although calculating $^{208}$Pb is no
different in principle from the triton, it is currently necessary to restrict
ourselves to a small (tractable) number of these coordinates.  Then and now
this restriction provides an operational definition of the few-nucleon systems:
those that we can accurately calculate.\\
\indent The first few-body conference that I attended was the 1974 meeting at
Laval University in Qu\'ebec City[3]. In my naivet\'e, I was both delighted and
disappointed.  People were actually trying to calculate the wave functions of
three- and four-nucleon systems from first principles, although it appeared
that they weren't doing a very good job of it.  Little did I realize how
complicated the problem was, nor that we had years to wait before computational
facilities became adequate to the scope of the problem.  Nevertheless, it was
this meeting that convinced me that accurate calculations and their comparison
with accurate experiments are the keys to the success of our field.\\
\indent In what follows I will comment on a few selected topics (concentrating
on interdisciplinary aspects) and try to contrast these observations with
similar material presented at Laval. Although most of the titles of the talks
presented there would not be out of place in this meeting, the level of
sophistication has increased greatly, and in some cases has a different
qualitative scale. I will also list in the form of ``bullets'' those
calculations and experiments that I believe need to be performed for orderly
progress in the field.\\
\indent Finally, at the end I will discuss a new approach that has already
provided insight into the nuclear force, and that might give much more. This
is the speculative part of the talk. It will be interesting to reread this in a
few years!

\begin{center}
THE FEW-NUCLEON SYSTEMS
\end{center}

\indent Traditionally, the 2-, 3-, and 4-nucleon systems comprised the
few-nucleon systems (FNS).  What makes their treatment so difficult (compared
to the atomic analogues) is the almost pathological nature of the
nucleon-nucleon (NN) force.  Not only are there very strong short-range
repulsive forces, but the most important part of the potential is the tensor
force, which mixes orbital and spin angular momentum.  In addition, there are
many (spin and isospin) components.  The recently developed Argonne V$_{18}$
force has 18 separate components, for example.  Because the Coulomb force is
much simpler than the nuclear one, our colleagues in atomic physics have
achieved a level of sophistication in calculations (which is exceeded only by
the precision of their experiments!) that truly boggles the mind.\\
\indent A certain class of few-nucleon calculations is often called ``exact''
or ``complete''.  My definition of such a class is that they solve the
Schr\"odinger equation for realistic nucleon-nucleon potentials with an
error of less than 1\%.  This was first achieved a decade ago for the triton
and we have added the $\alpha$-particle ($^4$He), the n-$\alpha$ resonances of
$^5$He and, recently, $^6$Li. These systems in my opinion correspond to the
\underline{current} definition of the few-nucleon systems. Table I shows the
unpublished Green's Function Monte Carlo (GFMC) results of Brian Pudliner[4]
for
$^3$H, $^3$He, $^4$He, and $^6$Li using the Argonne V$_{18}$ potential and the
Urbana model-9 three-nucleon force adjusted to reproduce the triton binding
energy.  Note that the theoretical results are the ones with error bars!
Clearly, there is no need for strong four-nucleon forces.

\begin{center}
Table I
\end{center}
\begin{tabbing}
xxxxxxxxxxxxxx\=xxxxxxxxxxx\=xxxxxxxxxxx\=xxxxxxxxxxx\=xxxxxxxxxxx\= \kill
\> {\underline{$^2$H}} \> {\underline{$^3$H}} \> {\underline{$^3$He}} \>

{\underline{$^4$He}} \> {\underline{$^6$Li}} \\
GFMC (MeV) \> 2.22 \> 8.47(2) \> 7.70(2) \> 28.35(8) \> 31.5(4) \\
Expt.\ (MeV)\> 2.22 \> 8.48 \> 7.72 \> 28.30 \> 32.0
\end{tabbing}

\indent An interesting program of variational calculations of light nuclei by
the Urbana-Argonne group[5] has made substantial progress in microscopic
calculations of $^{16}$O and $^{40}$Ca.   This technique makes use of a cluster
expansion, which appears to converge rather well, at least for $^{16}$O.  The
binding energy/nucleon that results is about 8 MeV, in good agreement with the
experimental value, for a potential that includes both the Argonne V$_{18}$ NN
force and a three-nucleon force (3NF) that has been adjusted to fit $^3$He.  It
would be very exciting if we could add all nuclei through $^{40}$Ca to our
list.  Although variational calculations have greatly improved in recent years,

$\bullet$ a marriage of GFMC and cluster techniques (if possible) would be a
great advance for our field.

\begin{center}
SOME RECENT PROGRESS
\end{center}

\indent Looking back 20 years, I marvel that any theoretical progress was made
using the computational facilities that were available at that time.
Calculations typically produced a few numbers, which were difficult to place in
any context.  As the facilities have improved, we are now able to generate a
flood of numbers, which can be analyzed to provide insight into few-nucleon
physics.  My favorite numerical methods text is Richard Hamming's ``Numerical
Methods for Scientists and Engineers''[6], which has the following epigraph:
``The purpose of computing is insight, not numbers.''  This clear and simple
directive is no surprise to a physicist, but without lots of numbers we can
have little insight.\\
\indent As an example of this, a ``scaling'' plot[7] of the charge radii of
$^3$H and $^3$He is shown in Fig. 1.  Each symbol without an error bar is a
calculation.
\vspace*{4.0in}

\begin{center}

FIG. 1. Scaling plot for rms charge radii of $^3$H and $^3$He vs.\ binding
energy.\\

\end{center}

The abscissa is the binding energy ($E_B$) of a particular nuclear force model,
which includes a Coulomb interaction for $^3$He, and in some cases a 3NF.  The
ordinate is the corresponding charge radius.  We see from this plot that a
realistic (i.e., physical) value for the latter can be obtained only if the
fitted curve is evaluated at the physical binding energy.  The average behavior
of the radius for $^3$He and $^3$H can be shown empirically to scale like
$1/E^{\frac{1}{2}}_B$, which follows immediately from the asymptotic behavior
of the wave function ($\sim e^{-\kappa \rho}$ with $\kappa \sim
E^{\frac{1}{2}}_B$).  We note that the divergence of the mean-square radius
with vanishing binding energy is very similar to the divergences with vanishing
pion mass that arise in chiral perturbation theory ($\chi$PT) calculations[8]
of the pion or nucleon charge (or magnetic) radii.  In the former case, the
nuclear center-of-mass (CM) is surrounded by a ``cloud'' of nucleons, while in
the latter case it is a cloud of pions.  The difference between the $^3$He and
$^3$H charge radii is due to the weaker force between two protons (in $^3$He)
or two neutrons (in $^3$H), and that between a neutron and a proton.  This
causes the two protons in $^3$He to lie further from the CM than the neutron
and conversely for the proton in $^3$H. Since the charge radius is
defined as the average distance of the charged particles (i.e., protons) from
the nuclear CM, the charge radius of $^3$He is greater than that of $^3$H.\\
\indent This geometric effect is therefore the result of the inequality of
forces between nucleons and is sometimes attributed to SU(4) symmetry breaking
or the S$^{\prime}$-state, which are different names for the same physics[9].
My
point is that if one had calculated only a few numbers and had been unaware of
the dependence of many observables on $E_B$, there could have been a serious
disagreement between theory and experiment.  In some capture reactions scaling
as severe as $E^{-2}_B$ has been observed[7].  Finally, we note that the same
mechanism leads to depolarization of the neutron in a polarized $^3$He
target[9], and to a nonvanishing neutron charge form factor in the naive
nonrelativistic quark model[10].\\
\indent Since this conference is a joint enterprise of nuclear and atomic
physics, I would like to comment on some very interesting recent experimental
work on the charge radii of the FNS. A contribution by the Yale group[11]
illustrated how the isotope shift between transitions in $^3$He and $^4$He can
be used to determine the \underline{difference} in rms radii of these systems.
Using the accurately known $^4$He result, the value for $^3$He is 1.9500(14)
fm. This number contains a lot of physics that we would like to test.  In order
for an electric field to probe a nucleus, it must first ``grab'' the charged
particles (the nucleons) before shaking the nucleus.  The nucleons themselves
have an intrinsic charge distribution that is folded with that of the wave
function. We will ignore other mechanisms, such as pion clouds, which can also
affect the nuclear size.  Removing the intrinsic nucleon radii leads[7] to
$$
<r^2>^{\frac{1}{2}}_{{\rm wfn}}\ = \hspace*{0.250in} 1.765(6) \hspace*{0.05in}
{\rm fm} \hspace*{0.750in} ({\rm expt.})\ , \eqno (1a)
$$
while extrapolating the upper (fitted) curve in Fig.\ 1 to $E_B(^3$He) =
7.72 MeV gives
$$
<r^2>^{\frac{1}{2}}_{{\rm wfn}}\ = \hspace*{0.250in} 1.769(5) \hspace*{0.05in}
{\rm fm} \hspace*{0.750in}({\rm theory})\ , \eqno (1b)
$$
where the theoretical ``error'' is a subjective estimate of the fluctuations
about the fit to the various calculations. The two experimental points in Fig.
1 were obtained from electron scattering, and the error on the $^3$He result is
seven times that of the Yale measurement. The agreement in Eqns. (1) is
excellent, but this highlights a long-standing problem with the deuteron.\\
\indent The deuteron rms radius extracted[12] from electron scattering data is
$$
<r^2>^{\frac{1}{2}}_{{\rm wfn}}\ = \hspace*{0.25in} 1.953(3) \hspace*{0.05in}
{\rm fm} \hspace*{0.750in}({\rm expt.})\ , \eqno (2a)
$$
while the recently constructed Nijmegen potentials[13] produce
$$
<r^2>^{\frac{1}{2}}_{{\rm wfn}}\ = \hspace*{0.25in} 1.968(1) \hspace*{0.05in}
{\rm fm} \hspace*{0.750in}({\rm theory})\ . \eqno (2b)
$$
Because many deuteron properties are determined primarily by its small binding
energy, and to a lesser extent by OPEP, deviations of even 1\% may point to a
breakdown of the impulse approximation, although calculations suggest
otherwise[12]. Atomic physics may have again come to the rescue.  A recent
determination of the isotope shift between hydrogen and deuterium in the 1S-2S
transition energy can be interpreted as a measurement of the deuteron radius if
other effects are removed.  This leads to[14]
$$
<r^2>^{\frac{1}{2}}_{{\rm wfn}}\ = \hspace*{0.25in} 1.973(7) \hspace*{0.05in}
{\rm fm} \hspace*{0.750in}({\rm expt.})\ , \eqno (2c)
$$
which is in much better agreement with the best theoretical value.  More
accurate experiments are under way.\\
\indent Finally, Fig. 1 shows that the tritium radius is not well determined.

$\bullet$ An accurate atomic measurement of the tritium-hydrogen isotope shift
would be extremely valuable.

\indent Two years ago I was asked to assess the status of theoretical
calculations in the three-nucleon systems[15].  My criterion was whether or not
complete calculations had been performed.  This typically requires that all NN
potential partial waves up to and including J = 3 be kept.  It is convenient to
break the 3N problem into four energy regions:  (1) bound states; (2)
zero-energy scattering; (3) scattering below the threshold ($E_{th}$) for
deuteron breakup in N-d scattering; and (4) scattering above that threshold.  I
further divided these problems into (a) incorporating NN forces only; (b)
including a 3NF; and (c) including a Coulomb interaction in the $^3$He (or p-d)
case.  Based on these categories, I constructed Table II.

\begin{center}
Table II
\end{center}

\begin{tabbing}

xxxxxxx\=xxxxxxxxxxxxxxxxxxxx\=xxxxxxxxxxxxxxx\=xxxxxxxxxxxxxxx\=xxxxxxxxxxxxxx
x\=\kill
\> \> {\underline{NN}} \> {\underline{3NF}} \> {\underline{C}} \\
\> $E = -E_B$  \> $\times$ \>  $\; \times$ \> $\times$ \\
\> $E = 0$     \> $\times$ \> $\; \times$ \> $\times$ \\
\> $E < E_{th}$\> $\, \bullet$  \> $\; \, \bullet$  \> $\,\bullet$  \\
\> $E > E_{th}$\> $\times$ \> $\; \, \bullet$  \> $\,$--

\end{tabbing}

\noindent A ``$\times$'' indicates that a complete calculation existed then.  A
``$\bullet$'' indicates that such a calculation now exists (or is underway) but
didn't then, while a ``--'' indicates that no calculation exists.  A single
Coulomb calculation by the Mainz group[16] using a limited number of partial
waves has been performed, and this calculation shows significant ($\sim$ 20\%)
Coulomb corrections at modest energies, warning us that comparing p-d data with
n-d calculations can be dangerous! Most of the entries in this table have
occurred since 1990.

$\bullet$ A major goal for the field should be complete p-d calculations above
breakup threshold with the Coulomb interaction included.

\begin{center}
SCATTERING CALCULATIONS AND EXPERIMENTS\\
\end{center}

\indent Scattering calculations (elastic and breakup) are very difficult, much
more so than bound-state calculations.  There were few contributions at Laval
treating this topic.  This has been the area with the greatest improvement, and
one where theory is beginning to play a vital (and historically unusual) role.
Because complete three-body scattering calculations are now tractable (indeed,
they are approaching maturity), theorists can make \underline{absolute}
predictions, which experimentalists love to test.  Failures mean that the
dynamics is inadequate and requires better forces.  It may be possible to
``see'' three-nucleon forces in scattering reactions, and a large experimental
effort has been mounted to look for these effects. Most calculations agree with
experiments very well without 3NF; only a few disagree[17]. A very good sign
from recent calculations is that some scattering observables appear to be
fairly
sensitive to these forces, although no definitive conclusions have been
reached.  Perhaps by the next meeting we will have a ``smoking gun''.\\
\indent I have been asked by a number of atomic colleagues why we place so much
emphasis on the Faddeev approach in our calculations, when this is not
necessary
in atomic physics.  The answer is ``boundary conditions''.  In situations where
the boundary conditions are self-evident, such as bound states or scattering
below the (energy) threshold for breakup (ejection of a previously bound
particle), any technique works well.  The extremely successful GFMC technique
is based on the Schr\"odinger, rather than the Faddeev, equation. Above breakup
threshold the boundary conditions are complicated and the Faddeev
decomposition, which leads to the Faddeev equations, is very effective in
implementing them.  Other techniques exist, but have been little used in
nuclear physics.  The Faddeev decomposition has been so successful as a
numerical procedure that it is also used for systems such as bound states where
the boundary conditions are obvious.\\
\indent One of the biggest advances in our field has been the increasing use of
polarized beams and targets by experimentalists.  The complexity of the nuclear
force requires detailed polarization information in order to disentangle the
dynamics.  Comparison of the contributions at Laval with those here highlights
this enormous improvement in capability.  I would guess that the majority of
the few-nucleon experiments reported here involved polarization. For one recent
example of how far this can be pushed, I refer the audience to the
Wisconsin[18]
p-d experiments at 3.0 MeV (lab).  With minimal assumptions two dozen phase
shifts and mixing parameters were determined. One of my biggest (and most
pleasant) recent surprises has been the successful implementation of the
hyperspherical-harmonic-expansion method for nuclei by the Pisa group[19].  By
incorporating explicitly the short-range correlations and using brute-force
numerical techniques this group has performed complete calculations of bound
states, zero-energy scattering (both n-d and p-d, with and without a Coulomb
force or a 3NF), and scattering below breakup threshold.  The bound-state
results agree with many other calculations, while the zero-energy scattering
agrees with our own calculations[15].  I can personally attest that the latter
were extremely difficult, so I am delighted that they have been confirmed.  The
Pisa results at finite energies appear to be in rather good agreement with the
Wisconsin analysis, except for partial waves sensitive to a 3NF, which can make
an appreciable difference. Coulomb effects were also shown to be important. I
hope that these groups get together and combine their talents.  Other
experiments at very low energies are eagerly awaited from TUNL.\\
\indent Repeating my earlier comment, implementing complete Coulomb
calculations for p-d scattering above breakup threshold remains a priority for
the field.  There are many very precise p-d experiments, which require complete
Coulomb calculations for their analysis.  The recent progress in this direction
by the Mainz group shows a substantial Coulomb effect and significant
differences between p-d and n-d scattering at modest energies.  Although these
very difficult calculations are not complete, they point to the danger of
comparing p-d data with n-d calculations.

\begin{center}
RELATIVITY AND NUCLEAR FORCES\\
\end{center}

\indent There were few contributions at Laval treating this subject, but many
here.  This is a somewhat controversial subject, but I will try to summarize
what has been done.  Naive estimates of $(v/c)^2$ in FNS can be obtained by
using the uncertainty principle.  We have $(v/c)^2 \sim (\bar{p}/M c)^2 \sim
(\frac{\hbar c}{M c^2 R})^2 \sim 1-2\%$, for radii, $R \sim$ 1.5 - 2.0 fm. What
this argument fails to account for is that relativity introduces operators that
depend on the nucleon momenta.  Expectation values of these operators probe the
tails of momentum distributions and are almost always larger than the estimate
above.  A number of calculations find that corrections to the kinetic and
potential energies are roughly 5\%, and because the NN system must be
reproduced these cancel to a large extent, changing the triton binding energy
by roughly 5\% . It is a measure of our recent success that we now need to
worry about effects of this size.\\
\indent About a year ago, I collaborated with the Nijmegen and Iowa groups[13]
to benchmark the triton binding energy for local potentials.  The Nijmegen
group
have been engaged in a sophisticated and extensive partial-wave analysis of NN
data.  They constructed a number of NN forces (updated Argonne, RSC, and
several varieties of Nijmegen potentials) in various versions.  Collecting all
of these, one finds that the local ones produced triton binding energies of
7.62(3) MeV, while momentum dependence of a particular type (from relativity)
produced \underline{additional} binding ($\sim$ 100 keV).  Other known
relativistic corrections have not been included in most potentials.\\
\indent The momentum-space Bonn B potential \underline{does} includes such
effects in part.  The Bonn potentials have always produced more triton binding
than any other potential, and the reason for this has been something of a
mystery.  Part of the reason is that the $^1S_0$ potential is a bit too strong
since it was fit to the n-p phase shifts.  Correcting for this, the triton
binding is still roughly 300 keV higher than the local potential result. In
contributions to this conference[20,21], it was shown that the choice of
$\pi$-N
coupling in the Bonn potential leads to a nonlocal OPEP (i.e., tensor force) of
relativistic origin, and this is presumably the reason for the difference. I
used the word ``choice'' because chiral-symmetry techniques show that other
forms are equally valid, and this choice is just one of many valid off-shell
extensions of OPEP. I view this improved understanding as a major success of
this meeting.\\
\indent We also need to ask whether it is really necessary to define
relativistic ``corrections''.  Why not simply use fully relativistic kinematics
and forces?  In principle the latter is the better approach, but current
practice dictates the former approach in most cases.  We still do not have
fully
satisfactory relativistic three-nucleon calculations.  Calculations of this
type are underway[22], but are not yet available.  For more massive nuclei they
are currently out of the question.  Semi-relativistic calculations (employing
$\sqrt{p^2 + m^2}$ for the kinetic energy of a nucleon) are feasible, however,
for such systems by using Carlson's clever trick[23] for implementing the
relativistic kinetic energy in configuration space.  Hopefully, modifications
of the potential (at least to order $(v/c)^2$) can also be added to this
procedure.

$\bullet$ Fully relativistic three-nucleon calculations will be required in
order to benchmark the reliability of semirelativistic calculations, and as an
accomplishment in their own right.

\indent There is another reason to follow this path.  Potentials have recently
been designed that provide an excellent fit to the NN data[13].  Triton
calculations seem to be indifferent to the details of the local potentials. On
the other hand, arbitrary (weak) nonlocalities can be added to the potentials,
which can still fit the data equally well.  We cannot use the NN system to
determine nonlocalities; we must use theoretical guidance, which highlights an
old problem:  we have a fundamental understanding of only a part of the nuclear
force. \\
\indent Fortunately, recent developments have eased this problem a bit.  The
Nijmegen group[24] have experimentally verified the presence of OPEP in the NN
force.  They fit the mass of the exchanged pions and find 139.4(10) MeV for the
charged pions and 135.6(13) MeV for the neutral ones, in excellent agreement
with the free masses.  The tiny error bars illustrate the importance of OPEP.
Indeed, OPEP produces about 70-80\% of the triton potential energy[13]. Given
this dominance, an obvious (and hopefully adequate) improvement to NN
potentials
would be to include relativistic corrections to OPEP and TPEP (two-pion
exchange), and these should be both well-founded and largest. Theorists in the
past had relatively few potential models to work with, and most of these had
known (and often irritating) defects. The more well-crafted models that we have
for our use, the better. We need to use potential models, not be their
prisoner.

$\bullet$ The development of semirelativistic potential models is needed.

\indent Ideally I would like to see an ``error bar'' of uncertainty
established, which is a measure of our ignorance for the FNS. For example, is
it possible to construct potentials with sufficient theoretical certainty that
the uncertainty in the triton binding energy ($\Delta$E) is constrained (e.g.,
$\Delta$E $\leq$ 0.25 MeV)? This would then translate into an uncertainty for
the 3NF.  Even though such a limit is probably the best we can do (without
directly solving QCD), it would be a wonderful accomplishment.\\
\indent Finally, a comment by Nathan Isgur[25] in his talk reminded me of an
obscure QED theorem, which in some sense explains why the hydrogen atom is so
simple if one doesn't look at the fine details. Nathan stated that the few-body
problem is scale dependent. That is, a change in the length or mass scale
changes the number of degrees of freedom (d.o.f.), or bodies, and hence this
conference depends on a choice of scale! He remarked parenthetically that even
the hydrogen atom is a many-body problem. Why? QED mandates that photons be
exchanged between the electron and the proton in all possible numbers and
orderings. If one were to take a snapshot of the atom at a particular time,
there would be photons in the air, making the atom a complicated many-body
problem involving the (virtual) photon ``glue''. Fortunately (or even
obviously), it was shown many years ago[26] that as one particle (the proton)
becomes very heavy, the electron acts as if it were moving in a static Coulomb
potential. In other words, the ``glue'' decreases as the $(v/c)^2$ of the heavy
particle decreases, and the problem becomes more potential-like. The same thing
can be expected to happen in a nucleus: the more nonrelativistic the nucleus,
the more potential-like the dynamics becomes.

\begin{center}
QCD AND ALL THAT\\
\end{center}

I became a physicist during a time when we had little idea about the
underpinnings of the strong interactions.  This was the era of ``you cannot use
perturbation theory because the couplings are too large'', or ``the model
uncertainties are too large''.  During the sixties one important event
changed things forever:  the discovery of chiral symmetry.\\
\indent Chiral symmetry resolved a major nuclear physics practical problem,
while providing explanations for other phenomena.  The Goldstone mode
realization of (broken) chiral symmetry makes the pion nearly massless on the
hadronic scale, and allows pion exchange to dominate the nuclear force under
ordinary conditions.  The major practical problem that was resolved was the
necessity of {\it ad hoc} ``Z-graph'' or ``pair'' suppression in relativistic
treatments of the nuclear force.  The off-shell pion-nucleon interaction can be
incredibly strong for arbitrary models, but is almost optimally weak for models
that obey chiral symmetry.  This use of chiral symmetry, which mandates that
multi-pion-exchange potentials (including 3NF) get progressively weaker, was
first proposed by Gerry Brown and collaborators[27]. Without this constraint
the nuclear force would be intractably complicated.\\
\indent  The second major influence of chiral symmetry was its incorporation as
an essential ingredient in QCD.  It is difficult now to imagine nuclear physics
without QCD.  Everyone ``talks'' QCD; ``finding'' quarks in nuclei has become
our search for the Holy Grail of medieval legend.  Unfortunately, this has had
minimal impact on our understanding of low-energy nuclear physics. The problem
is that quarks (constituents) and gluons (the binding mechanism) don't exist as
free particles.  Moreover, it only requires collisions between nuclei with a
few MeV of energy to experimentally demonstrate that \underline{nucleons} are
the most important degree of freedom in a nucleus (at any modest energy scale).
More energy ($\sim$ 150 MeV) shows that pions are the next most important,
etc.  How then can we interpolate between these two descriptions of nuclei,
which appear so fundamentally different, even contradictory?\\
\indent I have spent a lot of time lately thinking about these issues, and
talking to experts in the field, who hold widely divergent views.  My own view
is summarized as:  degrees of freedom in physics are a choice, not an
obligation.  If one chooses to describe a system with ``good'' d.o.f., the
subsequent description will be economical and ``clean''.  Choosing ``bad'' ones
will be uneconomical and ``ugly''.  Neither set is correct or incorrect. A
possible way out of this dilemma has been afforded by Weinberg[28] (and many
others, including Gasser and Leutwyler[8]).  QCD can be ``mapped'' onto a set
of
``effective'' d.o.f., which might, for example, include nucleons, pions, and
$\Delta$ isobars.  A field theory for the interactions between them can then be
constructed that manifests the correct (broken) chiral symmetry, and
potentials can be constructed using this interaction.  OPEP is an immediate and
not unexpected result.  This scheme is called Chiral Perturbation Theory, and
is a surrogate for the underlying QCD.\\
\indent Unfortunately, this scenario is not expected to work well above an
energy scale, $\Lambda$, on the order of 1 GeV.  Above that energy there should
be a gradual transition to another (and more economical) description.
Consequently, all short-range (very massive, M $\rsim \Lambda$) objects except
for the nucleon and possibly the $\Delta$ are
considered more appropriate to the underlying QCD and are ``frozen out''.
Because massive particles cannot propagate very far, their effect is replaced
by $\delta$-functions with arbitrary coefficients.\\
\indent My first impression of $\chi$PT was that it had a good ``feel'' to it,
because it combined principle and phenomenology in a way that nuclear
physicists have always done (or tried to do).  Phenomenology is required
because the $\delta$-function coefficients that subsume the influence of hard
(short-range) QCD effects must be fit to data.  This scheme is exemplified by
the meson sector, where more than 10 observables are neatly fit by only two
parameters[29]. There have been successes in the nucleon sector, as well.  It
should be noted that this field theory is not renormalizable in the old sense
of that term, but this is not a problem, as we were told by Peter Lepage[30].\\
\indent To lowest order, the NN potential is given by OPEP plus
$\delta$-functions that represent $\rho$-, $\omega$-, ... exchanges. Recall
from quantum mechanics that the $\delta$-function potential gives a sensible
\underline{constant} Born approximation scattering amplitude, but diverges in
higher orders.  There are two ways to handle the divergences. One is to fatten
the $\delta$-function (i.e., make it have finite
extent[31]) and the other is to renormalize the perturbation expansion in some
way. Although the latter approach would be the cleaner alternative, I have no
idea how to do it in an economical way.  Our traditional nuclear physics
methods
are economical, if they are nothing else.  For example, we ``renormalize'' by
using form factor cutoffs.  Finding a way to resolve this technical problem, or
showing that we cannot do it at all, is in my view extremely important.

$\bullet$ In my opinion one of the major theoretical problems in our field is
to determine whether $\chi$PT can provide a sound and economical framework for
understanding few-nucleon systems.

\begin{center}
ARE NUCLEI SOFT AND NATURAL?\\
\end{center}

\indent Not to be confused with an advertisement for shampoo, a positive answer
to the question could make a profound impact on nuclear physics. The question
is
another way of phrasing the ``bullet'' above.  For any dynamical scheme to work
effectively, it needs to converge rapidly.  This is especially true for schemes
based on $\chi$PT, since each order of perturbation theory introduces more and
more unknown constants, which can only be determined from experiments. Clearly,
if convergence is not rapid there is little hope for this approach.\\
\indent In nuclear physics, there are basically two ways to proceed with
dynamics.  We can either directly calculate amplitudes from the
underlying fields (the ``QHD'' approach[32]), or one can calculate a potential
from these ingredients and calculate amplitudes using the potentials.  The
potential scheme is the one we are most familiar with in treating FNS.  It is
rather efficient because, after a strong, hard interaction between two
nucleons and before another collision occurs, the nucleons can ``coast''.  In
weakly bound systems (or in scattering) the Green's functions responsible for
``coasting'' can be nearly singular, and this infrared singularity means that
successive collisions are (roughly) as important as the first one.
Consequently, a very complicated amplitude can be obtained (in practice) from a
much simpler (in principle) object:  the potential (providing that we can
calculate the latter).\\
\indent  As stated above, the dynamics of $\chi$PT is expressed as a series in
dimensionless variables[33,34]. The two mass scales of the problem are
$\Lambda$($\sim$ 1 GeV), the QCD large-mass scale, and $f_{\pi}$ ($\sim$ 93
MeV), the pion decay constant. For energy or momentum scales less than (or
comparable to) $\Lambda$, nucleons, pions, etc., should be a good set of
d.o.f.\ to use.  We heard an entire session of talks that suggests that this is
reasonable. At a recent CEBAF workshop we learned from the practitioners that
in
electromagnetic interactions with nuclei for increasing momentum transfers, $q
> \Lambda$, any and all ``improvements'' in the physics give roughly the same
effect, implying to me that in this regime mesons and nucleons may be
increasingly poor choices for d.o.f.\\
\indent The Lagrangian of $\chi$PT can be written as a series in the various
(unspecified) fields with these length scales and dimensionless
coefficients[34]:
$$
{\cal L} \sim \sum^{}_{i,j} c_{ij} \left[\frac{()}{\Lambda}\right]^i
\left[\frac{()}{f_{\pi}}\right]^j \ . \eqno (3)
$$
If nuclei are ``natural'', then $c_{ij}$ = Order(1), and if nuclei are
``soft'',
the series converges in some sense.  As an example[31] of ``natural'', strong
interaction coupling constants should have the form
$$
\frac{G}{\Lambda} \sim \frac{1}{f_{\pi}} \ , \eqno (4)
$$
or $G \sim 10$.  Thus, large strong-interaction coupling constants are
natural.\\
\indent Following Weinberg[28] and van Kolck[31], we can use power counting in
$\Lambda^{-n}$ to estimate the size of various contributions to potentials,
which leads to
$$
\left< V_{NN} \right> > \left< V_{3N} \right> > \left< V_{4N}\right>
\cdot \cdot \cdot \  . \eqno (5)
$$
This result, that successive multi-nucleon forces get progressively weaker, had
already been anticipated at the beginning of the talk.  In the triton, 3NF
cannot contribute more than $\sim$2\% of NN forces, while 4NF appear not to be
needed at all in the $\alpha$-particle or $^6$Li. Recently, van Kolck[31] has
used the same ideas to demonstrate that charge dependence in the NN force is
bigger than normal charge-symmetry breaking (n-n vs.\ p-p), which is bigger
than CSB in the n-p system:
$$
\left< V_{CD} \right> > \left< V_{CSB} (T = 1) \right> > \left< V_{CSB}
({\rm n-p}) \right> \ , \eqno (6)
$$
which holds experimentally, but was always something of a mystery to me.\\
\indent The approach to nuclear dynamics that we have described here is very
unusual, and it remains to be seen how well it works in practice.  I believe
that it has promise beyond the dominance proofs in Eqns. (5) and
(6). I will finish by quoting Steve Weinberg[35]:  ``...the chiral Lagrangian
approach turns out to justify approximations (such as assuming the dominance of
two-body interactions) that have been used for many years by nuclear
physicists...''. Hopefully, there will be more successes.

\begin{center}
ACKNOWLEDGEMENTS\\
\end{center}

This work was performed under the auspices of the U.\ S.\ Department of Energy.
\pagebreak

\begin{center}
REFERENCES\\
\end{center}

\begin{enumerate}
\itemsep=-0.02in
\parsep=0.0in

\item C.\ Wilkin, Lectures given at 1967 C.\ A.\ P.\ Summer School, Montreal,
BNL Preprint 11722.

\item J.\ L.\ Friar, Nucl.\ Phys.\ {\bf A156}, 43 (1970); M.\ Fabre de la
Ripelle, Fizika {\bf 4}, 1 (1972).

\item Few Body Problems in Nuclear and Particle Physics, Proceedings of
International Conference held at Laval University, Qu\'ebec City, Canada, 1974,
ed.\ by R.\ J.\ Slobodrian, B.\ Cujec, and K. Ramavataram (Les Presses de
l'Universit\'e Laval, Qu\'ebec, 1975).

\item Brian Pudliner and Vijay Pandharipande, Private Communication.

\item S.\ C.\ Pieper, R.\ B.\ Wiringa, and V.\ R.\ Pandharipande,
Phys.\ Rev.\ C {\bf 46}, 1741 (1992).

\item R.\ W.\ Hamming, Numerical Methods for Scientists and Engineers,
(Dover, New York, 1986).

\item J.\ L.\ Friar, Czech.\ J.\ Phys.\ {\bf 43}, 259 (1993).

\item J.\ Gasser and H.\ Leutwyler, Ann.\ Phys.\ (N.\ Y.\ )
{\bf 158}, 142 (1984); J.\ Gasser, M.\ E.\ Sainio, and A.\ \v Svarc,
Nucl.\ Phys.\ {\bf B307}, 779 (1988).

\item J.\ L.\ Friar, in Proceedings of Workshop on Electronuclear Physics with
Internal Targets and the BLAST Detector,  ed.\ by R.\ Alarcon and M.\ Butler,
(World-Scientific, Singapore, 1993), p.\ 210.

\item J.\ L.\ Friar, Particles and Nuclei {\bf 4}, 153 (1972); F.\ Klein,
Invited talk presented at this conference.

\item D.\ Shiner, R.\ Dixson, and V.\ Vedantham, Contributed paper at this
conference, p.\ 699.

\item S.\ Klarsfeld, J.\ Martorell, J.\ A.\ Oteo, M.\ Nishimura, and
D.\ W.\ L.\ Sprung, Nucl.\ Phys.\ {\bf A456}, 373 (1986).

\item J.\ L.\ Friar, G.\ L.\ Payne, V.\ G.\ J.\ Stoks, and J.\ J.\ de Swart,
Phys.\ Lett.\ {\bf B311}, 4 (1993).

\item K.\ Pachucki, M.\ Weitz, and T.\ W.\ H\"ansch,
Phys.\ Rev.\ A {\bf 49}, 2255 (1994).

\item J.\ L.\ Friar, in Proceedings of Workshop on Electronuclear Physics with
Internal Targets and the BLAST Detector,  ed.\ by R.\ Alarcon and M.\ Butler,
(World-Scientific, Singapore, 1993), p.\ 230.

\item E.\ O.\ Alt and M.\ Rauh, Contributed paper at this conference, p.\ 15.

\item W.\ Gl\"ockle, Invited talk presented at this conference.

\item L.\ D.\ Knutson, L.\ O.\ Lamm, and J.\ E.\ McAninch,
Phys.\ Rev.\ Lett.\ {\bf 71}, 3762 (1993).

\item A.\ Kievsky, M.\ Viviani, and S.\ Rosati, Pisa Preprint IFUP/TH 15/94.

\item J.\ Adam, Invited talk presented at this conference.

\item Y.\ Song and R.\ Machleidt, Contributed paper at this conference,
p.\ 189; A.\ Amghar and B.\ Desplanques, Contributed paper at this conference,
p.\ 547.

\item A.\ Stadler and F.\ Gross, Contributed paper at this conference, p.\ 922.

\item J.\ Carlson, V.\ R.\ Pandharipande, and R.\ Schiavilla,
Phys.\ Rev.\ C {\bf 47}, 484 (1993).

\item R.\ A.\ M.\ Klomp, V.\ G.\ J.\ Stoks, and J.\ J.\ de Swart,
Phys.\ Rev.\ C {\bf 44}, R1258 (1991); V.\ Stoks, R.\ Timmermans,
and J.\ J.\ de Swart, Phys.\ Rev.\ C {\bf 47}, 512 (1993).

\item N.\ Isgur, Invited talk presented at this conference.

\item S.\ Deser, Phys.\ Rev.\ {\bf 99}, 325 (1955).

\item G.\ E.\ Brown, A.\ M.\ Green, and W.\ J.\ Gerace, Nucl.\ Phys.\
{\bf A115}, 435 (1968); G.\ E.\ Brown and J.\ W.\ Durso, Phys.\ Lett.\
{\bf 35B}, 120 (1971).

\item S.\ Weinberg, Nucl.\ Phys.\ {\bf B363}, 3 (1991);
Phys.\ Lett.\ {\bf B251}, 288 (1990);
Phys.\ Lett.\ {\bf B295}, 114 (1992).

\item J.\ F.\ Donoghue, E.\ Golowich, and B.\ R.\ Holstein, Dynamics of the
Standard Model, (University Press, Cambridge, 1992); S.\ Scherer and H.\ W.\
Fearing, Contributed paper at this conference, p.\ 918.

\item G.\ P.\ Lepage, Invited talk presented at this conference.

\item C.\ Ord\'o\~nez and U.\ van Kolck, Phys.\ Lett.\ {\bf B291}, 459 (1992);
C.\ Ord\'o\~nez, L.\ Ray, and U.\ van Kolck, Phys.\ Rev.\ Lett.\ {\bf 72},
1982 (1994);
U.\ van Kolck, Thesis, University of Texas, (1993);
U.\ van Kolck, Contributed paper at this conference, p.\ 930.

\item B.\ D.\ Serot, Repts.\ on Prog.\ in Physics {\bf 55}, 1855 (1992).

\item A.\ Manohar and H.\ Georgi, Nucl.\ Phys.\ {\bf B234}, 189 (1984).

\item B.\ W.\ Lynn, Nucl.\ Phys.\ {\bf B402}, 281 (1993).

\item S.\ Weinberg, in Proceedings of the XXVI International Conference on High
Energy Physics, Dallas, 1992, ed.\ by J.\ R.\ Sanford, AIP Conference
Proceedings {\bf 272}, 346 (1993).

\end{enumerate}
\end{document}